\documentclass[11pt]{article}
\usepackage{graphicx,wrapfig}
 
\begin{document}
 
\title{Faster Monte Carlo Simulations at Low Temperatures. The Waiting Time Method}
\author{Jesper Dall and Paolo Sibani\\  
Fysisk Institut, SDU--Odense Universitet\\
 Campusvej 55, DK--5230 Odense M\\
 Denmark
} 
\date{\today}
\maketitle
 
\begin{abstract} 
\noindent We discuss  a  rejectionless global optimization technique 
 which, while being  technically similar to 
the recently introduced method of Extremal 
Optimization,   still relies on a physical analogy with
a thermalizing system.
This   method can be  used at constant temperature
or combined with annealing 
techniques, and is   especially well suited 
for studying the low temperature relaxation
of complex  systems as glasses and  spin glasses.

\noindent {\bf }
%\vspace{3.5cm}
\noindent {\bf }
\end{abstract}

\section{Introduction}  
The archetype of optimization approaches 
 inspired by physics is arguably Simulated 
Annealing (SA)~\cite{Kirkpatrick83,Cerny85}. 
In SA,     putative solutions of  the problem 
at hand  are  organized in an `energy landscape' similar to
the configuration space of a physical system.
This  landscape is searched  for the global optimum 
using the   Metropolis  algorithm~\cite{Metropolis53} 
while the temperature  is   slowly decreased
during the simulation. 
The theoretical underpinning of SA is equilibrium statistical physics
 and the theory of Markov processes: at each 
temperature $T$,   the
Metropolis algorithm  asymptotically   
samples  the equilibrium Boltzmann distribution  which,
at $T=0$,    is supported on the
set of ground states. 

It is well known to practitioners of SA  that      
maintaining thermalization  at arbitrarily   low 
temperatures is practically impossible 
in landscapes  with many local minima:
  The trajectories get trapped in subsets of 
configuration space (valleys or pockets) surrounding local minima,
and most of the attempted moves are rejected. Thus, ergodicity
is broken, and the final solution  is   just 
a `good' local optimum,  not  the global optimum.  

To mitigate  the effects of  ergodicity breaking  
one must  make sure  that   many different valleys are 
thoroughly explored. Ways to do so
is to employ ensemble methods~\cite{Ruppeiner91} or
use non-monotonic temperature schedules, as done
in bouncing~\cite{Schneider98}, thermal cycling~\cite{Mobius97}, 
and tempering~\cite{marinari98:sg}.  
Another possibility   is to select candidate moves from 
a broad distribution~\cite{Szu87,Tsallis96},   occasionally 
producing  the  `long jumps' in configuration space which 
are needed to  escape the neighborhood of a local minimum.

In this paper  we discuss  and test   a 
`thermal'  Monte Carlo scheme, which we have chosen to
call the  \emph{Waiting Time Method} or WTM. 
The WTM belongs to a familiy of rejectionless
or `event driven' algorithms~\cite{Binder79} 
which developed  from the  so-called n-fold method 
introduced by  Bortz et al.~\cite{Bortz75}.
From an \emph{algorithmic} point of view, the WTM
is  very close  to  a   recent 
optimization method, 
Extremal Optimization (EO)~\cite{Boettcher00,Boettcher00a},
notwithstanding  the fact that the latter is
inspired  by   the 
Bak-Sneppen model of biological evolution~\cite{Bak93}, rather
than thermal annealing of a physical system. 
 
Rejectionless  methods can offer
much improved performance, especially
in simulations of glassy systems at
low temperature.  Yet, they  
  do not seem to have gained
widespread acceptance~\cite{LandauBinder00}.
We  hope that the  present analysis  
will contribute  to  improve the
situation.

As a prelude, we   show the equivalence of the WTM
with the Metropolis algorithm    with regard  to the
equilibrium  as well as to
the dynamical (relaxation) properties.
Secondly, we analyze the performance of the WTM  and
identify the temperature region where it  is
computationally advantageous to use it.  
Finally we compare the performance of  
Extremal Optimization with 
the WTM run at constant temperature and with the WTM in combination with
the Geman-Geman annealing schedule~\cite{Geman84,Hajek88}. The numerical
analyses  all use a 3D Ising spin glass model with short
 range Gaussian interactions as a test problem.  
 
\section{The WTM algorithm}

The WTM is well suited for   problems  where 
 $N$ variables 
contribute to the cost function (energy)
 in an additive fashion. One such case  is
 the short range Ising spin glass model:
$N$ spins, $\sigma_i = \pm 1$  are  
placed on a  cubic lattice, producing a total of 
$2^N$ configurations.  
Each configuration has $N$  neighbors, which correspond  to
the $N$ possible  flips of a single spin.
 The      energy   of configuration $\alpha$ is given by
\begin{equation}
{\cal H} = -\frac{1}{2}\sum_i \sigma_i^\alpha \sum_j J_{ij} \sigma_j^\alpha,
\label{energy_def}
\end{equation}
where   the couplings $J_{ij}$ are 
independently 
drawn from a Gaussian distribution
 of unit variance and zero mean if $i$ and $j$ point  to
adjacent lattice sites. Otherwise $J_{ij}=0$.

For any configuration\footnote{We will often leave the dependence on 
$\alpha$ understood.} $\alpha$, the WTM  associates to each
  spin  a stochastic variable  $t_i$ which
is the time  at which this  spin   must flip, given that its  local field 
$\sum_j J_{ij} \sigma_j$   stays put.
A  `global time' value, which is
initially set to zero,  is  stored in the variable $t_{\rm global}$. 

Let $\Delta_i$ be the energy change associated  with the flip 
of $\sigma_i$.
Flipping  times $t_i$ are initially assigned as
\begin{equation}
t_i = -\tau_{i} \log X_i,
\label{waitingtimes}
\end{equation} 

\noindent where  the $X_i$  
 are   independent  stochastic  variables  with a uniform distribution in the
unit interval. Thus, each $t_i$ has an exponential probability distribution 
with average waiting time given by
\begin{equation}
\tau_{i} =  \max(1,\exp(\Delta_i/T)).
\label{tau_i}
\end{equation} 

\noindent After the initialization  has been completed, 
the WTM   iterates the following three steps:
\begin{enumerate}
\item Flip the spin $\sigma_{m}$ having the lowest flipping time.
\item Update the global time: $t_{\rm global} =   t_{\rm m}$.
\item Generate from an exponential distribution 
with average $\tau_i$ fresh   waiting times $\delta_i$  for
the spin $\sigma_{m}$ and its lattice neighbors 
(e.g. six other spins in a cubic lattice). 
The waiting times  are added    to the global time  
in order to obtain  the updated  
flipping times: $t_i = t_{\rm global} + \delta_i $. 
\end{enumerate}
 In order  to identify the  next move, one needs to 
search  (and update) a  database of size $N$ 
 storing  the current flipping time
of  each degree of freedom   as well as  pointers to the $z$ 
 degrees of freedom  interacting with it. 
Ordering in a balanced tree (a heap),
 with the shortest flipping time  
placed at the root node,  requires a  
computational effort   for the search and update 
procedure of order $\log N$. Since the  
number of updates  per spin flip
is equal to  $z+1$,
the computational overhead per spin flip  is 
$(z+1) \log N$, independent of the temperature $T$.

\section{Theoretical considerations}
For completeness, we discuss below  a few 
basic  statistical properties   of the WTM.
The main results are derived analytically
and then checked numerically for
illustration purposes. We also briefly
describe Extremal Optimization, which
is used in the performance comparison of
the next section.

\subsubsection*{Equilibrium distribution for the WTM}
It is   simple to  establish that the 
WTM fulfills  \emph{detailed balance}~\cite{Binder79}. 
Consider two configurations,  $\alpha$ and $\beta_k$, the 
latter obtained from the former by flipping spin $k$.
Given that  spin $k$ is the next spin to flip, the probability 
that the system remains in state  $\alpha$  for at least  a 
 time $t$ is  (by construction) 
\begin{equation}
q_k(t) = \exp(-t/\tau_k).
\end{equation}
The rate $R_{\beta_k,\alpha}$ of probability flow 
from $\alpha$ to $\beta_k$ is then
\begin{equation}
R_{\beta_k,\alpha} = \frac{1}{\tau_k} =  \min(1,\exp(-\Delta_k/T)).
\label{probflowrate}
\end{equation}
Detailed balance follows since   
$R_{\beta_k,\alpha}/R_{\alpha,\beta_k} = \exp(-\Delta_k/T) =
 P_{\rm eq}(\beta_k)/P_{\rm eq}(\alpha)$, 
where $P_{\rm eq}(\alpha)$ is the probability
 of visiting configuration $\alpha$ in equilibrium.
As a consequence hereof~\cite{vanKampen92}, the
stationary distribution of the WTM is the  Boltzmann
distribution over the   states of the system,  
as  in the case of the  Metropolis  algorithm.

\subsubsection*{Metropolis and   WTM dynamics}\label{MET&WTM}
Since detailed balance is fulfilled,
the   WTM has  the same equilibrium
(Boltzmann) distribution as the Metropolis algorithm.
To an excellent 
approximation,   there is  also a 
 \emph{dynamical} equivalence, in the sense that the 
 transition probabilities from one state to 
 another are very nearly  
 the same\footnote{If a   number  of waiting times $\tau_i$, 
 which all equal  one,  happens to enter the heap in close succession,
 the WTM  has a bias  towards  flipping  the corresponding spins
 in the same   order as the $\tau_i$ are submitted.
 This feature hardly has any  practical consequences and is
 neglected  in the   formulae.}
 in the two schemes. 
This  equivalence 
simplifies a performance comparison considerably,
since one only needs to compare the computer
times spent on average to achieve a flip. 

The WTM remains in a  state $\alpha$ 
for   a time at least  equal to
  $t$   with a  `survival' probability given by 
\begin{equation}
Q_\alpha^{\rm W}(t)  = \prod_{k=1}^N \exp(-t /\tau_k) = 
\exp(-t \sum_{k=1}^N \tau^{-1}_k). 
\label{waittime1}
\end{equation} 
The  probability  that spin $k$ be the first to flip is 
\begin{equation}
p_k^{\rm W} = \int_0^\infty 
{\tau_k^{-1}}Q_\alpha^{\rm W}(t)dt
= \frac{\tau_k^{-1}}{\sum_j \tau_j^{-1}}. 
\label{flip-first}
\end{equation}
In the   Metropolis algorithm, the 
 transition probability $P(\beta_k,\alpha)$ 
 from a configuration $\alpha$ to one of its $N$
neighbors $\beta_k$  is  
\begin{equation}
P(\beta_k,\alpha) =  \frac{1}{N} \min(1,\exp(-\Delta_k/T)).
\label{metro}
\end{equation}
Hence,   the probability
 that   spin $k$  be  the first to flip is
\begin{equation}
p_k^{\rm M} =   \frac{  P(\beta_k,\alpha)}{\sum_j P(\beta_j,\alpha) }.
\label{norej}
\end{equation}
Since Eq.~(\ref{flip-first}) is identical to Eq.~(\ref{norej}), the
WTM and the Metropolis method generate  the same Markov chain
of spin flips.

We also note that the intrinsic   (or `physical') time of the WTM  
approximately  corresponds to the number of MC   
steps  in the Metropolis algorithm.
The  latter  leaves   state $\alpha$  unchanged 
for at least  $t$  MC  steps with 
the probability 
\begin{equation}  
Q_\alpha^{\rm M}(t) =
\left(
 1 - \sum_{k=1}^N  P(\beta_k,\alpha) 
\right)^{tN}. 
\label{waittime2}
\end{equation} 
If $\alpha$ is a local minimum  all the $\Delta_k$  are positive and 
the sum  in Eq.~(\ref{waittime2}) is close to zero for small $T$.
A similar conclusion applies to configurations close to a local minimum,
which is where the algorithm will spend  most of the computer time
in a multi-minima landscape at low $T$.
In this situation  one easily obtains  
\begin{equation}
Q_\alpha^{\rm M}(t)\approx \exp[-t \sum_{k=1}^N \min(1,\exp(-\Delta_k/T)  )].
\label{waittime3}
\end{equation}
Hence, using the usual Monte Carlo step as time unit 
makes Eq.~(\ref{waittime1}) and~(\ref{waittime3}) 
identical. Note however    that Eq.~(\ref{waittime3})
is an approximation while  Eq.~(\ref{waittime1}) is exact.

\begin{figure}[t]
\begin{center}
\includegraphics[width=9.5cm]{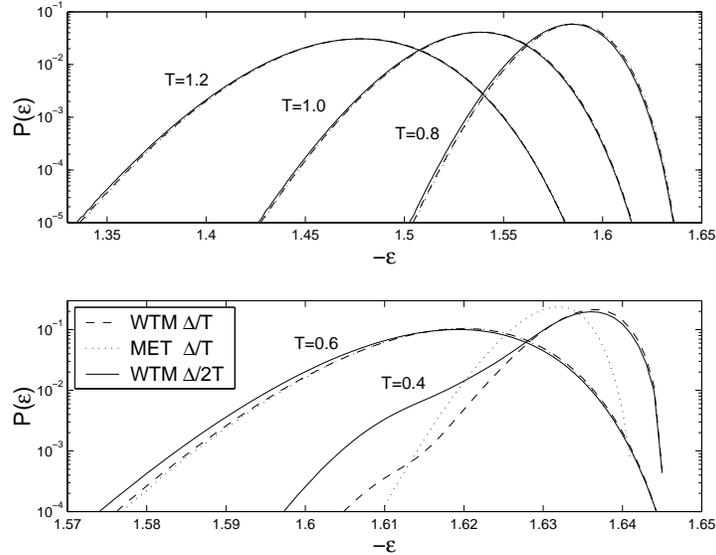}
\caption{\small The normalized
density of states  $P(\varepsilon)$ for a
Gaussian Ising spin glass   with $N=8^3$ spins 
is  sampled  through $10^6$ MC steps. 
The  Metropolis algorithm and two variants
of the   WTM, which are based upon 
Eq.~(\ref{tau_i}) and Eq.~(\ref{tau_i_2T}), 
respectively, were utilized to obtain the data.
See the main text for further details.  }
 \label{MTvsWTM}
\end{center}
\vspace{-0.5cm}
\end{figure}
Fig.~\ref{MTvsWTM} shows, for
 different temperatures,  the
density of states of the spin glass problem sampled with
the WTM (dashed lines), the Metropolis algorithm (dotted
lines) and finally the WTM with a different   choice of rates, 
(henceforth "choice b") given in Eq.~\ref{tau_i_2T} and  
discussed below. 
In all cases we
used $10^6$ MC steps: At $T=1.2$ and $T=1.0$ this  
suffices  to equilibrate the system (the  
discrepancies barely visible at high energies are more
pronounced for smaller systems (not shown) and are likely to stem from 
a finite size effect). At $T=0.8$ 
 the WTM and Metropolis are still
indistinguishable. The incipient lack
of equilibration transpires from the fact that  
 the data sampled by the WTM with  choice b of rates 
now visibly differs from the other two sets at
high energies. This  difference is enhanced at 
  $T=0.6$, while  at $T=0.4$ the trajectories
 become  trapped  in different local minima and 
the three methods   strongly differ due to  insufficient 
statistics.

\subsubsection*{Other choices of   the average waiting time}
Within all possible choices of rates fulfilling
detailed balance,  
Eq.~(\ref{probflowrate}) stands out as the one 
which, besides producing the desired
equilibrium distribution, also   guarantees  
dynamical equivalence  with  the Metropolis algorithm.      
In systems like spin glasses,  where the relaxation   
is dominated by partial equilibrations in   
configuration space traps, one nevertheless   
expects  the dynamics to be   insensitive 
to the precise choice of rates. 
To  check  this, we  considered the choice
\begin{equation}
\tau_{i} =  \exp(\Delta_i/2T),
\label{tau_i_2T}
\end{equation} 
whose appeal mainly lies in the
fact that  the same expression  for
$\tau_i$ applies 
regardless of the  sign of  $\Delta_i$. 
 
\begin{figure}[t]
\begin{center}
\includegraphics[width=9.5cm]{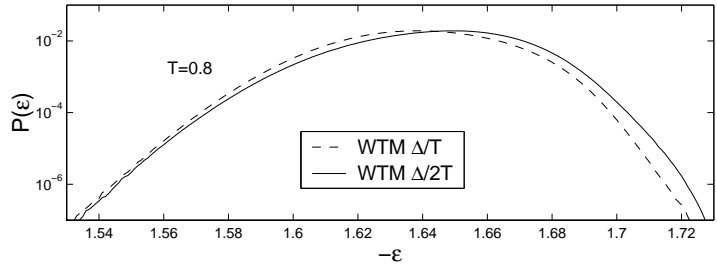}
\caption{\small $P(\varepsilon)$
averaged over $10^4$ runs on the same $N=8^3$ 
gaussian spin glass. All runs started at the
same  configuration with  $\varepsilon = -1.6200$ 
and  lasted only 100 Monte Carlo steps. 
The two versions of the WTM refer to Eq.~(\ref{tau_i}) 
and Eq.~(\ref{tau_i_2T}).}
\label{1&2WTM}
\end{center}
\vspace{-0.5cm}
\end{figure}
In Fig.~\ref{1&2WTM} we compare  the
densities  of states sampled at $T=0.8$ by the
 WTM with   rates chosen according to
Eq.~\ref{tau_i_2T} and Eq.~\ref{probflowrate}.
The latter choice is, we recall, equivalent to 
the Metropolis scheme. The data shown are averages over
$10000$ very \emph{short} runs. All these runs started from 
the same initial state, with energy per spin  $\varepsilon = -1.6200$,
and  lasted  $100$ MC steps, which is  not even 
remotely sufficient for equilibration purposes.
As anticipated, the   densities 
of states sampled by the two methods are
quite similar, even at very short time times.

\subsubsection*{Extremal Optimization}
The WTM is  in many respects  \emph{algorithmically} similar
 to Extremal Optimization~\cite{Boettcher00,Boettcher00a}.
 We have, unsuccessfully, tried 
to map the two methods  onto each other, and
we also failed  to produce an analytical description 
 of the stationary distribution belonging to EO. In short, 
with optimal choice of parameters for optimization,
%% $\gamma$ and $T$,  
$P_{\rm EO}(\varepsilon)$ is significantly broader 
than $P_{\rm WTM}(\varepsilon)$~\cite{Dall00}. 
  
The quality of the contribution to the
overall cost  from a  degree of freedom 
is considered in EO as a  \emph{fitness}
 measure. The actual way in which the fitness is assigned is not rigidly
specified, but reflects one's intuition and knowledge of the
problem at hand. In our case, we use  the
local energy contribution $-\sigma_i \sum_j J_{ij} \sigma_j$
as the fitness $f_i$ of $\sigma_i$, in accordance with~\cite{Boettcher00b}.
The following steps are then  iterated:
\begin{enumerate}
\item Rank the spins according to their fitness, and pick a 
spin with rank $n$ with probability $p \propto n^{-\gamma}$.
The exponent $\gamma$  is a free parameter which
can be tuned for a given class of problems  
 but which is not changed dynamically  during the runs.
\item Flip the chosen spin and calculate the ensuing change 
in the fitness of its  neighbors.
\end{enumerate}
If  $\gamma$ is very large,   the least fit spin  is chosen every time, and
EO resembles  the Bak-Sneppen model of evolution~\cite{Bak93}. 

The major   difference  between EO and WTM is that
only  the rank (relative  magnitude)
    of the energy changes is important in the former method, 
    while the absolute magnitudes are  
    important in  the latter. Secondly, EO 
 chooses its   next move probabilistically
from a deterministic ranking  of the $f_i$.
By way of contrast,
the  WTM chooses the next move   deterministically
from a list of waiting times   produced by 
a stochastic algorithm.  Thirdly,     
the ranking procedure in EO utilizes  the current energy
contribution, while the WTM looks ahead and 
generates the waiting times according to the 
projected energy change.

\section{Tests and results} \label{test&results}
This section  presents  the salient features of the
WTM algorithm, found by simulations of 
 the 3D Gaussian spin glass model with $N$ spins.
 We do not attempt a complete characterization,
which would be tedious for those not interested in the
particular model  used. Likewise, the physical aspects of
low temperature relaxation in spin glasses 
will be discussed elsewhere~\cite{DallSib01b}.

\subsubsection*{Acceptance rates and intrinsic WTM time}
\begin{figure}[t]
\begin{center}
\includegraphics[width=9.5cm]{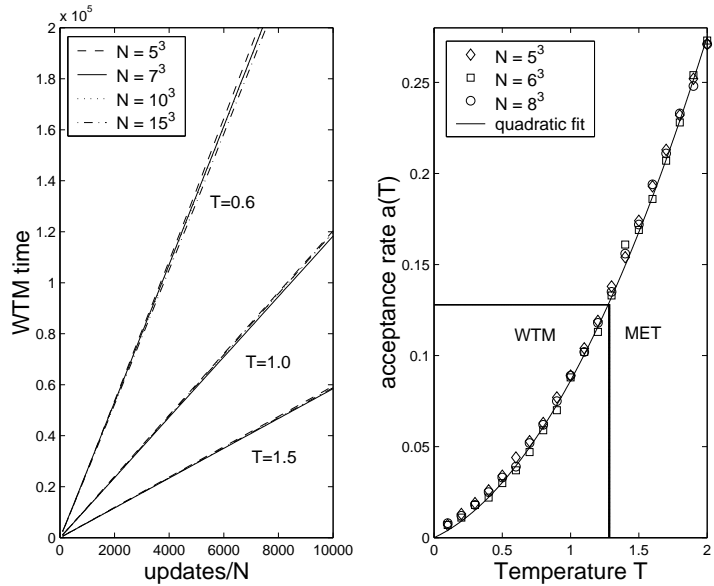}
\caption{\small The left panel shows the 
elapsed intrinsic time in the WTM as a function of the number of updates
measured in MC steps. Results for three different temperatures and a 
number of system sizes are displayed.
 The right panel shows the acceptance
rate, i.e the ratio of 
the number of flips to the number of
attempted  flips,  for the Metropolis algorithm, 
with the box marking the temperature
region --  $T < 1.3$ -- where the WTM is faster. }
 \label{WTM_portr}
\end{center}
\vspace{-0.5cm}
\end{figure}

The `intrinsic' or physical time elapsed in a WTM simulation
determines the amount of configuration space explored and 
how close the system is to thermal equilibrium.    
Its linear relation to the number of Monte Carlo steps
 is displayed in the left panel
of Fig.~\ref{WTM_portr} for three different
temperatures and a number of systems sizes. 
There is no discernible  $N$ dependence, except at the
lowest temperature, where the \emph{smallest} systems, $N=5^3$ and
$7^3$, are seen to follow slightly steeper lines 
(longer times)~\cite{DallSib01b}.

The  right panel of Fig.~\ref{WTM_portr} shows, for 
  three different system sizes,  
the empirical temperature dependence of the acceptance rate $a(T)$ 
of  the Metropolis algorithm, i.e the ratio of 
the number of flips to the number of attempted  flips. 
Again, the size dependence is negligible. 

The slopes in the left panel are, for each temperature, nearly equal 
to the reciprocal acceptance rate $1/a(T)$  obtained 
from the right panel.   This confirms   the 
 correspondence between the WTM time $t$   
and the number of MC steps discussed in 
  Section~\ref{MET&WTM}.
  
\subsubsection*{Tuning of parameters}
In   ergodic systems, and  in particular in  finite systems,
the best-so-far energy per spin
 $\varepsilon_{\rm bsf}(t) $ seen in  a simulation of length $t$ 
must approach  the ground state energy as $t \rightarrow \infty$,
regardless of the temperature. 
Nonetheless, for realistic values of $t$ and all but the
smallest systems, the ground state energy is never reached 
and one can meaningfully study the parameter dependence of 
a suitably averaged the best-so-far energy
$\langle \varepsilon_{\rm bsf}(t) \rangle$.

In applying the   WTM to the spin glass  problem, 
 we   averaged over many runs with the same length 
 and  different $J_{ij}$.  
We  observed  a  clear temperature dependence of the 
resulting $\langle \varepsilon_{\rm bsf} \rangle$ and defined
 an  `optimal'  parameter $T_{\rm opt}$   as the temperature minimizing 
$\langle \varepsilon_{\rm bsf} \rangle$
after $10^3 N$ spin flips. Empirically, this   
 $T_{\rm opt}$ is   \emph{independent}   of the runtime
through  several orders of magnitude~\cite{Dall00}.
It is however    dependent on the system size in the following way:
For $N =  5^3,6^3,8^3,10^3$ and $12^3$, we find  
 $T_{\rm opt} = 1.0,0.9,0.7,0.65$ and $0.6$,  respectively.

While we cannot make any
theoretical statements regarding EO, the behavior 
with respect to $\langle \varepsilon_{\rm bsf} \rangle$
is very much the same, with $\gamma$ in lieu of $T$.
However, running EO on the same systems as above showed a
much weaker $N$ dependence: though $d\gamma_{\rm opt}/dN<0$ 
is observed, $\gamma = 1.1$ produces good results
for $N\leq 12^3$. Compared to the temperature in the WTM, 
$\gamma_{\rm opt}$ is insensitive to changes 
in the system size.

\subsubsection*{Runtime considerations}
\begin{figure}[t]
\begin{center}
\includegraphics[width=9.5cm]{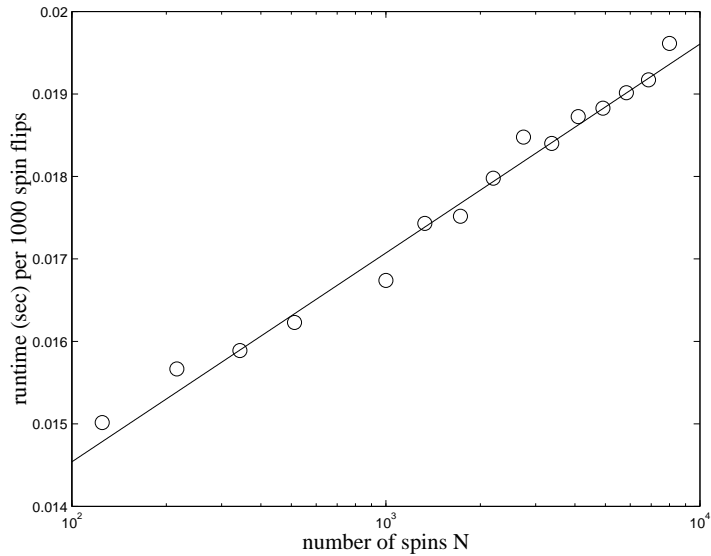}
\caption{\small WTM on a 500MHz Pentium III. 
The plot shows the logarithmic increase in runtime as a function of
 the system size $N$, due to the updating of the binary tree in which the
spins are stored according to their flipping times.
%Observe the slow increase in runtime due to the ordering: 
%The runtime for  $N=20^3$ is only about $30\%$ larger than for $N=5^3$.
}
 \label{WTM_runtime}
\end{center}
\vspace{-0.5cm}
\end{figure}

A ball park figure
for the  temperature $T_b$  where the WTM  is faster than  Metropolis
can be obtained by noticing that 
the generation of waiting times, see Eq.~(\ref{waitingtimes}),
 is the most time consuming
part of the algorithm.  For each performed spin flip
the WTM    generates   $z+1=7$ new waiting times, while 
 Metropolis requires $1/a(T)$ attempted spin flip operations on average.
 Equating the two expressions yields $a(T_b) \simeq 1/7 \approx  0.15$
 and $T_b \approx 1.5$. The actually measured  value
 of $T_b$ is indicated by the  lower corner of the box
 in  Fig.~\ref{WTM_portr} and equals $ \approx 1.3$. This is  somewhat
lower than the estimate, likely because we  neglected 
 the computational cost of order $(z+1)\log N$, 
confirmed in Fig.~\ref{WTM_runtime}, which is needed in  
the search and update procedure. From Fig.~\ref{WTM_portr} 
we see that at $T=0.5$  
the WTM is approximately three times 
as fast as standard Metropolis. A closer examination of $a(T)$ 
reveals that at $T \simeq 0.25$ and $T \simeq 0.035$ the WTM 
is 10 and 100 times faster, respectively, if both methods are
to produce the same number of spin flips. 

\subsubsection*{Performance comparisons}
\begin{figure}[t]
\begin{center}
\includegraphics[width=9.5cm]{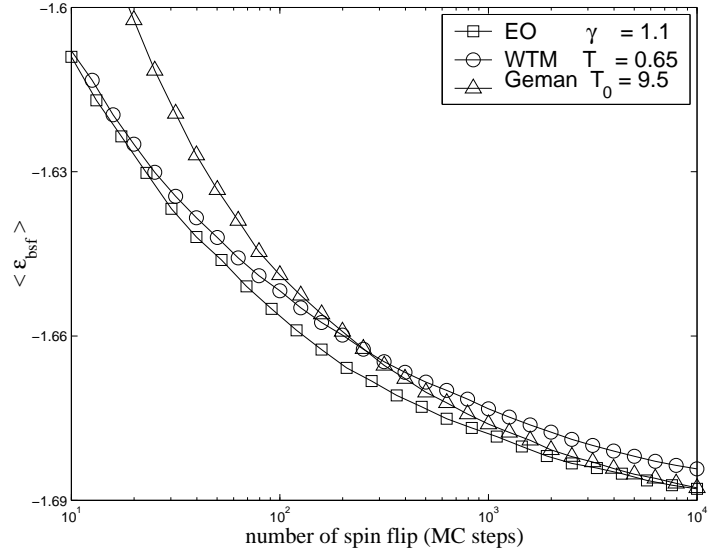}
\caption{\small A comparison of the energy per spin  
$\langle \varepsilon_{\rm bsf} \rangle $,
 averaged over runs on 200 different spin glasses,
for three different methods on a $N=10^3$ 
Gaussian spin glass. The  legend specifies  the 
EO exponent, the WTM temperature and the initial
 temperature for the Geman-Geman schedule.}
 \label{Compare}
\end{center}
\vspace{-0.5cm}
\end{figure}

We have compared the WTM
with an annealing version and with EO. 
The result is shown in Fig.~\ref{Compare}.
In the  annealing case we have used the Geman-Geman logarithmic schedule
$T(t) = T_0/(1+\log(1 + t))$.
We observed the reasonable fact that $\langle \varepsilon_{\rm bsf} \rangle$
is minimized when the initial temperature $T_0$  
is chosen so that the final
temperature is slightly below $T_{\rm opt}$ in the WTM.

The WTM is seen to be inferior to EO in terms of spin flips
needed to reach the same value of
 $\langle \varepsilon_{\rm bsf} \rangle$. However, when implementing
EO ad literam\footnote{which is usually not 
done~\cite{Boettcher00,Boettcher00a}, since an exact ordering is
costly and an approximate ordering is assumed to work just as well.},
at equal \textit{runtime} EO and the WTM produce similar results 
for $N=10^3$, i.e. they
reach approximately the same $\langle \varepsilon_{\rm bsf} \rangle$. 
For larger systems EO is slower than the WTM~\cite{Dall00}.

When applied to an 
annealing schedule like Geman-Geman, the WTM is seen to 
perform better than EO, since varying the temperature in
the WTM does not change the runtime notably.
We note in passing that while the Geman-Geman schedule is usually 
dismissed as unrealistically 
slow, it performs quite satisfactorily in connection with WTM.

%%\begin{Large}
%%\begin{center}
%%$\stackrel{\stackrel{\stackrel{\frown \quad\frown}
%%{\circ \quad \circ}}{\stackrel{\%%parallel}{{\smile \smile}}}}{\smile}$
%%\end{center} 
%%\end{Large}

\section{Summary and conclusions} 

The WTM belongs to a familiy of rejectionless algorithms
related to the \emph{n-fold way} of Ref.~\cite{Bortz75}.
These algorithms  asymptotically sample  the Boltzmann
distribution and are mathematically
equivalent to  the ubiquitous Metropolis algorithm.
Compared to Metropolis, they   also require a computational
and implementation  overhead. However they are considerably faster
at low temperatures, especially   in systems with short   
range interactions.
The  n-fold way, in particular,   is efficient  
if  $n$ is small, i.e. if the possible
values of the  $\Delta_k$ belong to  
a small, discrete set. 
The WTM method does not suffer from the same limitation, 
and  is therefore    especially    well suited for low 
$T$ simulations of the dynamics of disordered and glassy
systems. 

Our tests on a spin glass system show 
that Simulated Annealing with Geman-Geman type
of schedule is slightly faster Extremal Optimization,
a recent addition to the arsenal of optimization 
tools. This indicates that `thermal' schemes
remain  a viable approach to global  optimization.  
   
\section{Acknowledgement}

We would like to thank Stefan Boettcher for kindly sharing his work
and ideas with us. P.S. thanks the participants of the 1999 
Telluride Workshop on Energy Landscapes for valuable input
and Statens Naturvidenskabelige Forskningsr\aa d for financial
support.

\bibliography{meld}
\bibliographystyle{unsrt}
\end{document}